\documentclass[12pt,preprint]{aastex}

\def\lapp{\ifmmode\stackrel{<}{_{\sim}}\else$\stackrel{<}{_{\sim}}$\fi}
\def\gapp{\ifmmode\stackrel{>}{_{\sim}}\else$\stackrel{>}{_{\sim}}$\fi}

\begin{document}

\title{On the X-ray Spectra of Anomalous X-ray Pulsars and Soft Gamma Repeaters}

\author{
V. M. Kaspi\altaffilmark{1,2,3} \&
K. Boydstun\altaffilmark{1,2}
}

\altaffiltext{1}{Department of Physics, Rutherford Physics Building,
McGill University, 3600 University Street, Montreal, Quebec,
H3A 2T8, Canada}

\altaffiltext{2}{Department of Astronomy, California Institute of Technology, Pasadena, CA 91125}

\altaffiltext{3}{Lorne Trottier Chair; Canada Research Chair; vkaspi@physics.mcgill.ca}

\begin{abstract}
We revisit the apparent correlation between soft X-ray band photon
index and spin-down rate $\dot{\nu}$ previously reported for Anomalous
X-ray Pulsars (AXPs) and Soft Gamma Repeaters (SGRs) by \citet{mw01}.
Our analysis, improved thanks to new source discoveries, better spectral
parameter measurements in previously known sources, and the requirement
of source quiescence for parameter inclusion, shows evidence for the previously
noted trend, although with greater scatter.  This trend supports
the twisted magnetosphere model of magnetars although the scatter suggests
that factors other than $\dot{\nu}$ are also important.  We also note
possible correlations involving the spectra of AXPs and SGRs in the
hard X-ray band.  Specifically, the hard-band photon index shows a
possible correlation with inferred $\dot{\nu}$ and $B$, as does the
degree of spectral turnover.  If the former trend is correct, then the
hard-band photon index for AXP~1E~1048.1$-$5937 should be $\sim$0--1.
This may be testable with long integrations by {\it INTEGRAL}, or
by the upcoming focussing hard X-ray mission {\it NuSTAR}.
\end{abstract}


\section{Introduction}

Anomalous X-ray Pulsars (AXPs) were long thought of as ``soft'' X-ray
sources, because their X-ray spectra below 10~keV were observed to be
falling off sharply. {\it INTEGRAL} and {\it Rossi X-ray Timing Explorer
(RXTE)} observations of these objects, as well as of their close cousins,
the Soft Gamma Repeaters (SGRs), revealed that their spectra turn over
in the 10--20 keV range, such that in fact some AXPs are the hardest
known sources in the sky in the hard X-ray band \citep{mcl+04,khm04,mgmh05,
dhk+06,gmte06}. Here we define the ``soft'' X-ray band as 1--10~keV,
and the ``hard'' X-ray band as 20--80~keV.

Prior to the discovery of the hard X-ray emission, \citet{mw01} (hereafter MW01)
reported an apparent correlation between soft-band spectral
hardness and spin-down rate $\dot{\nu}$, where $\nu$ is the spin frequency.  
This reported correlation has been used
as the primary motivation for the ``twisted magnetosphere'' model for
magnetars \citep{tlk02,tb05,bt07}. In that model, surface activity driven
by the decay of the large internal magnetic field of the star results in
a twisted dipolar morphology of the magnetosphere, with powerful currents
emanating from the surface and subsequently returning to heat it.  In this
picture, the MW01 hardness/$\dot{\nu}$ correlation is interpreted
physically as a correlation between hardness and spin-down-inferred
surface dipolar magnetic field $B$, where the latter can be estimated
via $B\equiv 3.2\times 10^{19} (P \dot{P})^{1/2}$~G, with $P\equiv
1/\nu$ the pulse period.  The simple $\dot{\nu}$/B connection exists because, 
for AXPs and SGRs, there is
a far wider range in observed $\dot{\nu}$ than observed $\nu$.  In the twisted
magnetosphere model, this correlation is a result of stronger currents,
due to higher $B$ hence higher activity, resonantly upscattering seed
thermal photons from the hot surface to form the non-thermal component.
In addition to the apparent soft-band hardness/$B$ correlation reported by
MW01, support for this picture comes from observed correlations
between soft-band spectral index and flux in individual variable sources
\citep[e.g.][]{tmt+05,roz+05,cri+07,tgd+08}, since all else being equal,
higher fluxes are expected from larger twists, because of higher return
currents.

In this {\it Letter}, we revisit the correlation reported by MW01 in
light of three main developments in the field since their work.  First,
several new sources have been discovered since 2001 and can be used to
confirm the original claimed correlation, which was based on only seven objects.
Second, previously known sources have been observed in longer integrations
with higher sensitivity instruments, allowing better determinations of
their spectral parameters.  Third, since 2001, it has been recognized
that AXPs can exhibit great variability both in soft X-ray flux and spectrum,
particularly during outbursts \citep[e.g.][]{kgw+03,wkt+04,icd+07,tgd+08}.  
However, MW01
considered measured magnetar spectral parameters without concern for the
objects' outburst status at the time of measurement, thereby possibly
using data from objects in very different spectral states.

In addition, with the study of the hard-band emission from AXPs and
SGRs well underway, there now exist enough data for correlations
similar to that noted by MW01 to be considered.  Indeed \citet{gmte06}
noted qualitatively that in the hard band, AXPs are harder than SGRs, the
reverse of what is seen in the soft band.  Here we consider this point
in more detail, noting possible correlations between the hardness
of the hard-band spectra of AXPs and SGRs and $B$ (and $\dot{\nu}$), and
between magnitude of spectral change from the soft to hard bands, and $B$
(and $\dot{\nu}$).

\section{Compiling AXP and SGR Spectra}

We characterize AXP and SGR soft and hard X-ray
spectra phenomenologically.  Although significant effort has been
expended to do physically motivated broadband spectral modelling
\citep[e.g.][]{lg06,rtz+07,og07,ntz08,zrtn09}, this work is still progressing, with no
final word reached.  Moreover, our motivation here is to characterize all
sources in a systematic and uniform way.  Hence we choose to parametrize
soft-band spectra using the simple blackbody
plus power-law model, with $\Gamma_s$ being the soft-band photon index
(defined as $dN/dE \propto E^{-\Gamma_s}$).
Similarly, we parameterize the hard-band spectrum by a power law of
photon index $\Gamma_h$ (defined as for $\Gamma_s$).  We recognize that this parameterization is
flawed, both in the soft band where the true spectrum is likely to be better represented by
a Comptonized thermal model (such as those currently under development)
as required in the twisted magnetosphere model, but also more dramatically
in the mid-energy range where there is clearly
spectral curvature.  Nevertheless the parameterization is convenient, used ubiquitously
in the literature,
and does a reasonable job of characterizing the soft and hard-band
spectra currently available, given statistical uncertainties \citep[see
e.g.][]{khdc06}.  We further acknowledge that
the spectra, both soft \citep[e.g.][]{pkw+01} and
hard \citep[e.g.][]{dkh08}, are dependent on pulse phase, however
here we consider only phase-averaged spectral parameters.

In Table~1, we summarize the results reported in the literature for all
known non-transient magnetars for which there are such measurements.
Since MW01, several new magnetars have been discovered, four of which
have well measured spin and spectral parameters (CXOU~J010043.1$-$721134,
SGR~0526$-$66, 1E~1547.0$-$5408 and CXO~J164710.2$-$455216).  In addition,
in the soft X-ray band, SGRs and AXPs are now established to be highly
variable, such that $\Gamma_s$ can vary greatly.
In compiling Table~1, we took great care to include the $\Gamma_s$
(always for a blackbody plus power-law model) for objects when their
reported fluxes were at or near the lowest yet recorded.  This was done
to ensure the objects were in their quiescent state, with no temporary
enhancement that could be interpreted as a magnetospheric twist with
corresponding spectral hardening.  By contrast, MW01 averaged
together different published values of $\Gamma_s$, presumably under
the assumption that the values do not change, which we now know to
be incorrect.  In making our selections from the literature, 
where multiple quiescent spectral observations were available, we 
verified that the reported $\Gamma_s$ values were roughly consistent,
then selected that with the smallest reported uncertainty.
We excluded the so-called transient AXP XTE~J1810$-$197 which
shows no clear non-thermal component when in quiescence \citep{ghbb04}.

\section{Soft-Band Spectral Indexes}

\begin{figure}[t]
\plottwo{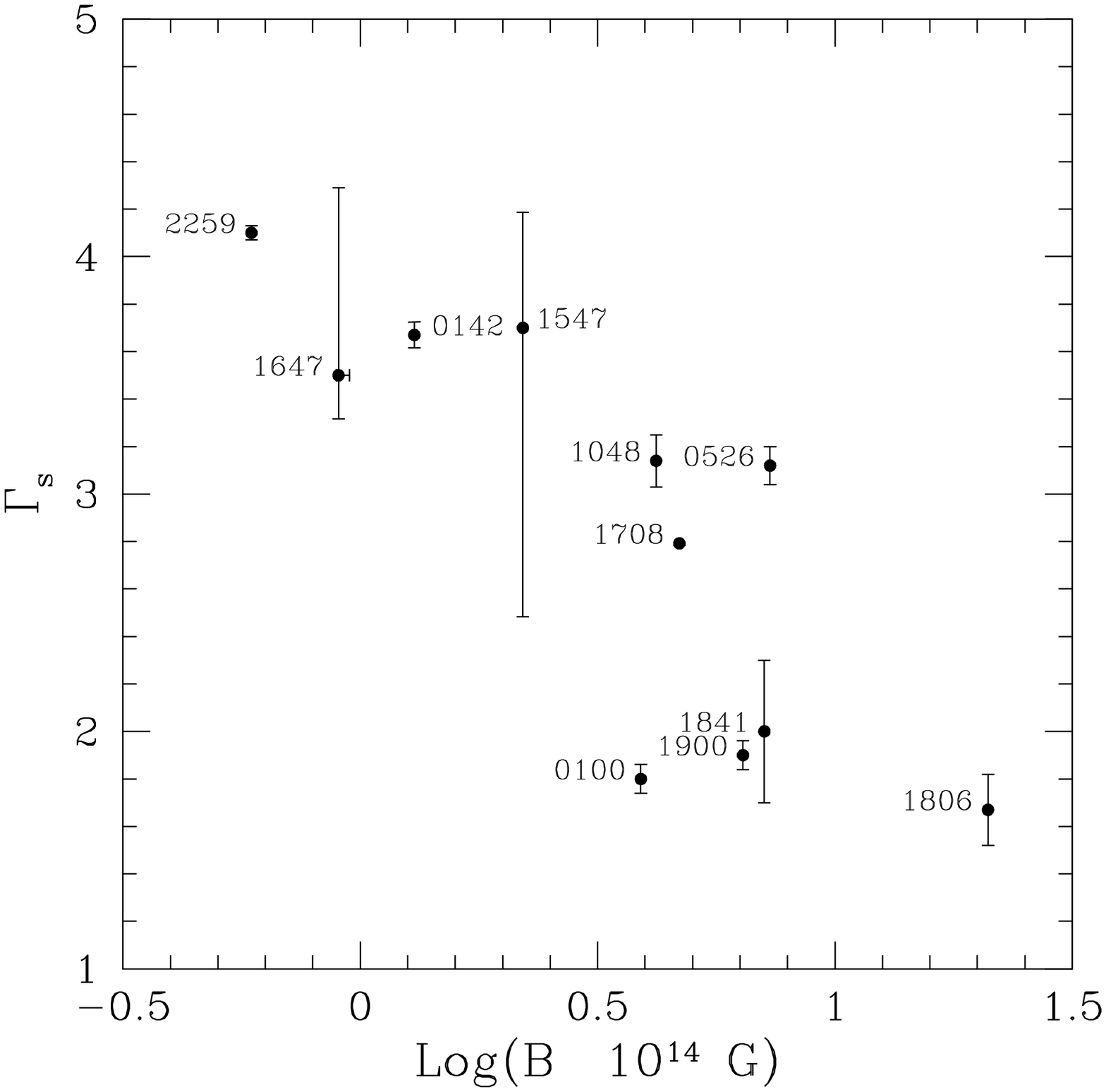}{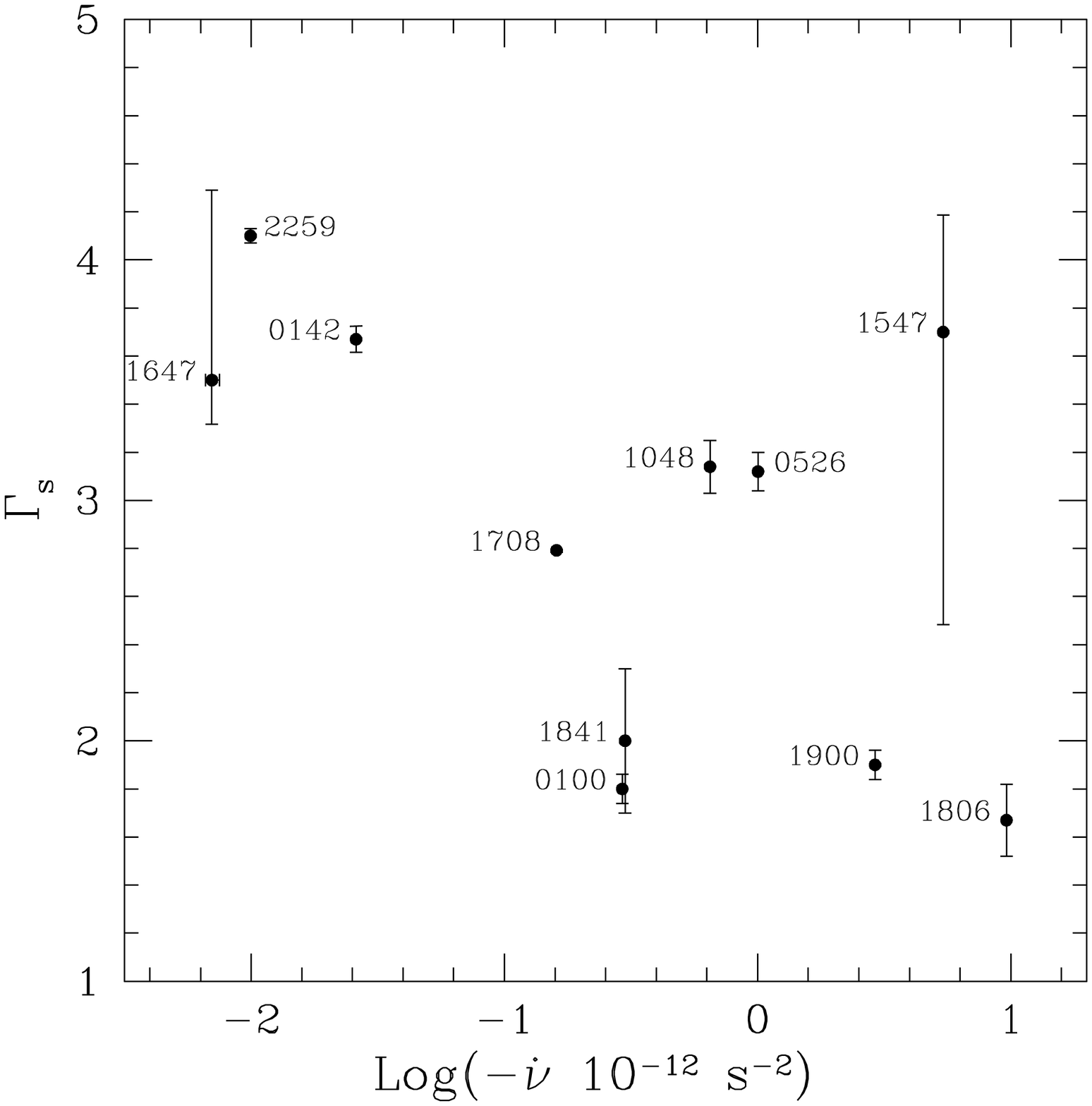}
\figcaption{
Left: $\Gamma_s$ versus $B$.  See
text and 
Table~\ref{ta:gammas} for data used and references.  
Right:  Same but versus $\dot{\nu}$.
Both:  Error bars represent 1$\sigma$ uncertainties.
\label{fig:gammas}
}
\end{figure}

Using the data provided in Table 1, we plot $\Gamma_s$ versus $B$ and
$\dot{\nu}$ in Figure~\ref{fig:gammas}.  Values of $\dot{\nu}$
are long-term averages as recorded in the SGR/AXP online
catalog\footnote{www.physics.mcgill.ca/$\sim$pulsar/magnetar/main.html} (see
references for $\dot{\nu}$ values therein).
We note the previously seen trend for decreasing $\Gamma_s$ with increasing $B$
or $\dot{\nu}$.  
To quantify this trend, we determined Pearson's linear correlation 
coefficient to be $r=-0.56$
for a sample size of $N=11$,
corresponding to a null-hypothesis (two-tailed) probability of $p=0.076$.
We note that this $p$ value, as for all others in this paper, is actually
an upper limit given that uncertainties are not accounted for in
calculating $r$.
This $\Gamma_s$/$\dot{\nu}$ trend is not
as striking as it was in MW01 (compare with their Fig. 2); there is
more scatter\footnote{We estimate for MW01's reported
$\Gamma_s$ vs. $\dot{\nu}$ correlation (their Fig. 2), $r=-0.83$ for $N=7$, implying $p=0.021$.}.  
If we ignore AXP 1E~1547.0$-$5408, whose
$\Gamma_s$ has large uncertainty, the trend is improved and is
significant with $>99$\% confidence ($r=-0.78$ for $N=10$, $p=0.0079$), though still with
considerable scatter.  Interestingly, the plot of $\Gamma_s$ versus $B$
(Fig.~\ref{fig:gammas}, left) shows a better correlation ($r=-0.82$
for $N=11$, $p=0.0022$), which
may support the twisted magnetosphere model in that it is the magnitude of
$B$ that determines the twist size, not just the spin-down rate.  Nevertheless,
there is scatter; this could imply that the value of $B$, as inferred from
spin-down parameters, is only a rough estimate of the true field strength,
a possibility already generally recognized given the greatly simplified
assumptions in computing $B$, and also observationally suggested from the
absence of anomalous X-ray emission from very high-$B$ rotation-powered
pulsars (e.g. Pivavoroff, Kaspi \& Camilo 2000; McLaughlin et al. 2003;
Kaspi \& McLaughlin 2005)\nocite{pkc00,msk+03,km05}.  

We also searched for a correlation between the simultaneously measured
blackbody temperature $kT$ and $B$ (see Fig. 2), but,
like MW01, found none.  Specifically, for $N=11$, $r=0.29$, implying
$p=0.38$, clearly non-excluding of the null hypothesis.  This argues
that the underlying physical parameter that is most closely linked with
a magnetar's magnetic field is manifested in the power-law component,
rather than in the thermal component \citep[see also][]{kkm+03}.

\section{Hard-Band Spectral Indexes}

Figure~\ref{fig:gammah} shows values of $\Gamma_h$ plotted versus $B$
(left) and $\dot{\nu}$ (right) for the six sources for which $\Gamma_h$
has presently been measured.  In some cases, $\Gamma_h$ is
for the pulsed emission, in other cases for the total emission,
and in three cases for both (see Table 1).
There is a possible trend in the data.
AXPs are harder in the 20--80~keV band than are SGRs as qualitatively
noted by \citet{gmte06}.  But as is clear here, even within the AXP
class itself, low $B$ or $\dot{\nu}$ sources are harder
than high $B$ or $\dot{\nu}$ sources.  
To be more quantitative, for the sources for which the pulsed emission
is measured, we find $r=0.88$ for $N=4$, implying $p=0.12$ for a $\Gamma_h$
correlation with $\log B$ (Fig. 3, left), and
$r=0.86$ for $N=4$ implying $p=0.14$ for the $\Gamma_h$ versus $\log \dot{\nu}$
data (Fig. 3, right).  These values are more significant for the
total flux ($r=0.79, N=6, p=0.062$ and $r=0.84, N=6, p=0.034$ for
$\log B$ and $\log \dot{\nu}$, respectively) when the (pulsed) value
for 1E~2259+586 is included, under the assumption that its total
flux $\Gamma_h$ does not differ much from its pulsed $\Gamma_h$, as
for the three sources for which both are measured.  Although not
quite significant at the $>99$\% confidence level, these possible
correlations seem noteworthy.

We note that the value of $\Gamma_h$ for 1E~2259+586 should be regarded
with caution, the source having been detected only up to 30~keV thus far.
This will hopefully improve with time.  We also of course caution that the
trend is defined by only six sources, one fewer than was used by MW01.
If this correlation is correct, then for AXP~1E~1048.1$-$5937, the
brightest AXP for which $\Gamma_h$ has not yet been measured, $\Gamma_h$
should be $\sim$0--1.  This may be testable with long integrations by {\it
INTEGRAL}, or by the future focussing hard X-ray mission {\it NuSTAR}
(expected launch 2011) which should have point-source sensitivity of
$\sim$100 times that of {\it INTEGRAL}.

Note that in all but two cases (4U~0142+61 and SGR~1806$-$20), the sources were
in quiescence at all epochs when the hard X-ray observations were done.
For 4U~0142+61, the measured hard X-ray spectral parameters are an average
over $\sim$4~yr of data \citep{dkh+08} which included a few-month active
period in 2006 \citep{gdk09}.  This active period
seems unlikely to have biased the hard X-ray spectrum because
of the absence of soft X-ray variability during that
epoch \citep{gdk09}, because of the relatively small duration of the
hard X-ray exposure during the active period relative to during
quiescence, as well as because of the absence of any evidence for
hard X-ray flux or spectral variability over the 4 years of {\it INTEGRAL} observations \citep[with flux and $\Gamma_h$ stable to within 17\%;][]{dkh+08}.  
For SGR 1806$-$20, the hard X-ray observations were
obtained roughly 2~yr after the source's giant flare of 2004 December.
During the hard X-ray observations, the 
source was nearly back to quiesence, though still bursting occasionally.
Whether this mild activity affected its hard X-ray spectrum is unknown
given that no variability has yet been reported in the hard X-ray spectrum
of {\it any} magnetar.  This will likely be difficult to determine with 
current hard X-ray telescopes given their sensitivity, barring another major outburst.

Figure~\ref{fig:turnover} shows values of $\Gamma_s - \Gamma_h$, our
parameterization of the degree of spectral turnover, plotted
versus $B$ (left) and $\dot{\nu}$ (right).  A trend is suggested by both
plots, such that the lowest
$B$ or $\dot{\nu}$ sources generally have the largest
spectral turnover.
Quantitatively, for the pulsed data, we find
$r=-0.99$ for $N=4$ implying $p=0.014$ and
$r=-0.98$ for $N=4$ implying $p=0.021$, for $\Gamma_s - \Gamma_h$ versus
$B$ and $\dot{\nu}$, respectively. 
When considering total flux, these values change to
$r=-0.90$ for $N=6$ implying $p=0.015$ and
$r=-0.92$ for $N=6$ implying $p=0.0087$, for $\Gamma_s - \Gamma_h$ versus
$B$ and $\dot{\nu}$, respectively, again including the (pulsed) data
for 1E~2259+586, under the reasonable assumption that the difference
between its total and pulsed flux spectral turnovers is small (as it
is for the three sources for which both have been measured).
Thus, these correlations are significant at the $\gapp 99$\% level.
If the $\Gamma_h$ versus $B$ trend holds and the above prediction
for $\Gamma_h$ for 1E~1048.1$-$5937 is correct, its spectral turnover
would agree well with this second trend.  The same obvious
caution regarding the paucity
of sources as in Figure~\ref{fig:gammah}, holds, however.

We also investigated whether there exists any correlation between hard
X-ray luminosity and $B$, but found none.  However, the luminosities of
the sources in this band are generally poorly known, given relatively
large $\Gamma_h$ and distance uncertainties.  Thus some correlation may
exist, but cannot presently be discerned.

\section{Discussion}

We have revisited the correlation between soft X-ray band power-law
spectral index and $\dot{\nu}$ and $B$ reported previously by MW01,
and find it continues to hold, though with significant scatter.  This stands
in contrast to the absence of any apparent relationship between blackbody
temperature and $B$ or $\dot{\nu}$.  This supports the foundation for
the ``twisted magnetosphere'' model of \citet{tlk02}, as further developed
by \citet{tb05} and \citet{bt07}.  
The significant
scatter in the still small measured population suggests in the context
of this model that the inferred $B$
is only a rough estimate of the stellar dipolar magnetic field, and/or
that other factors, such as geometry (both intrinsic and viewing) and
higher-order magnetic multipoles, play physically important roles.
With very recent discoveries of several new magnetars, for example SGR
0501+4516 \citep{enr+09} and SGR 0418+5729 \citep{hck+09}, as well as
new measurements of spin parameters for previously known objects, as,
for example, for SGR 1627$-$41 \citep{ebp+09}, it should be possible to
further test the veracity of the correlation in the near future.

The possible correlations we have noted for the
hard-band emission from magnetars may be telling us something new
about the magnetar emission process.  Next we consider whether the
published models have predicted the observed trends.

\citet{hh05} suggested that magnetohydrodynamic fast modes produced
by magnetars may break down to form electron/positron pairs in the
magnetosphere.  They suggested that non-thermal radiation associated
with this fast-mode breakdown could account for the high-energy emission
from magnetars.  However in this model, the spectrum of the radiation
is fixed and is not expected to vary from source to source as has been
observed.  This difficulty, along with the now falsified \citep{dkh+08}
prediction that the fast-mode breakdown radiation spectrum should extend
above 1~MeV, argues against this model being relevant to explain the
observed properties of the hard X-ray emission in magnetars.

\citet{bh07} explore 
resonant inverse Compton upscattering of thermal emission
from the surface, by relativistic electrons only a few stellar radii
from the surface.  This resonant scattering, at least in
their preliminary analysis, is expected to yield a very flat spectrum,
independent of $B$, unlike what is
reported here.  However, \citet{bh07} did not consider non-resonant
scattering which they suggest may have a very different spectrum,
although its relative contribution is as yet unclear.  For the resonant
scattering, although substantial spectral differences are expected for
different scattering locales, as well as for more complicated field
geometries than the assumed dipole, the flatness of the
spectrum independent of $B$ seems robust.
The correlations we report thus suggest that this process
is not the dominant mechanism for producing the hard X-ray emission
in magnetars.

\citet{tlk02}, \citet{tb05} and \citet{bt07} argue that powerful
magnetic currents in the magnetar magnetosphere are induced by sporadic
starquakes that twist the external magnetic field. A self-induction
electric field lifts particles from the surface, accelerates them, and
yields avalanches of pair creation.  This forms a corona close to the
neutron star surface. \citet{tb05} and \citet{bt07} consider a model in
which the hard X-rays are produced in the transition layer between the
corona and thermal photosphere.  In this model, the hard X-ray emission
is a result of thermal bremsstrahlung with a temperature that scales
approximately as $kT \propto B^{2/5}$, with $kT \simeq 100$~keV for
reasonable parameters.  However, we verified, using spectral simulations,
that this scaling predicts the opposite of the correlation seen in
Figure~\ref{fig:gammah}, because higher $kT$, expected for higher $B$,
gets fit with a lower $\Gamma_h$ when fit with a simple power law, for
$kT$ in the range 20--200~keV.  Thus the observed correlation does not
seem consistent with a thermal bremsstrahlung origin.

On the other hand, the observed correlation between the spectral index
change and $B$ could be interpreted at least qualitatively in the
framework of models in which magnetospheric currents play a role in the
production of the high-energy emission as follows.  In the magnetosphere,
relativistic electrons can resonantly scatter surface thermal photons
to form the non-thermal portion of the soft-band spectrum as discussed
first by \citet{tlk02}.  Alternatively, electrons that do not have a
chance to scatter will impact on the surface, heating it, and possibly
creating a corona as described by \citet{tb05} and further explored
by \citet{bt07} \citep[but see][]{le07}.  For higher $B$ sources, the
scattering optical depth increases, rendering the soft index harder as
described by \citet{tlk02}, but leaving fewer electrons for heating the
corona, and hence possibly resulting in the hard index being softer.

For energetic rotation-powered pulsars (RPPs),
the power-law spectrum seen below 10~keV extends well beyond
that energy, such that the spectral turnover
$\Gamma_s - \Gamma_h \approx 0$ for practically all sources
for which it has been measured.  In a handful
of cases, there is marginal evidence for
a subtle change of shape of the hard spectrum, 
generally parameterized as a log-parabolic
steepening above $10$~keV \citep[e.g.][]{cmd+08}.
Clearly if the trend in Figure~\ref{fig:turnover} held
for RPPs, one would expect huge turnovers,
given their far lower $B$.   This argues
strongly that the mechanism for the production of hard X-rays in magnetars
is different in RPPs, which was already suggested by the fact that the
magnetar hard X-ray emission
has luminosity far in excess of that available from rotational
slow-down, in contrast to that from RPPs.  
$\Gamma_h$ for the usually rotation-powered
PSR~J1846$-$0258 \citep[but see][]{ggg+08}, the highest-$B$ ($4.9\times
10^{13}$~G) RPP for which $\Gamma_h$ has been
measured \citep{kh09}, is 1.2, much larger than would be predicted by
the trend in Figure~3, left ($\Gamma_h$ versus $B$) but smaller than what
would be predicted in Figure~3, right ($\Gamma_h$ versus $\dot{\nu}$).
Indeed the same is true of practically all RPPs for which $\Gamma_h$ has
been measured, further demonstrating a different hard-X-ray production
mechanism.

We thank A. Beloborodov, D. Eichler, M. Livingstone, and C. Thompson for useful
conversations.  We thank Caltech for its
hospitality and support via a Summer Undergraduate Research Fellowship
(KB) and a Moore Scholarship (VMK).  VMK receives additional support
from NSERC, FQRNT, CIFAR, and holds a Canada Research Chair and
the Lorne Trottier Chair in Astrophysics and Cosmology.


\newpage

\newcommand{\marka}{\tablenotemark{a}}
\newcommand{\markb}{\tablenotemark{b}}
\newcommand{\markc}{\tablenotemark{c}}
\newcommand{\markd}{\tablenotemark{d}}
\begin{table}[t]
\begin{center}
\caption{Spectral Parameters for Non-transient SGRs and AXPs\marka}
\scriptsize{
\begin{tabular}{cccccccccc} \hline\hline
Source Name 			& $B$ & $\dot{\nu}$ 		& BB $kT$ & $\Gamma_s$ 			& Ref.  & $\Gamma_h^p$ & Ref. & $\Gamma_h^t$ & Ref. \\
            			& ($10^{14}$ G)	& ($10^{-12}$ s$^{-2}$) & (keV)				&				& for $\Gamma_s,kT$	&				& for $\Gamma_h^p$	&		& for $\Gamma_h^t$ \\\hline
1E 2259+586 			& 0.59   & -0.0099 		& 0.411(4) 			& 4.10(3) 			& 1 	& $-$1.02$_{0.13}^{0.24}$	& 12 	& ... 		& ... \\
CXO J164710.2$-$455216		& $\sim$0.9\markb& $-$0.007\markb & 0.49(10)\markc		& 3.5$_{0.3}^{1.3}$\markc 	& 2 	& ...  				& ... 	& ... 		& ... \\
4U 0142+61  			& 1.3 & $-$0.026 		& 0.386(5)\markc		& 3.67(9)\markc			& 3 	& 0.40(15) 			& 13 	& 0.93(6) 	& 13 \\
1E 1547.0$-$5408 		& 2.2 & $-$5.41		& 0.43$_{0.04}^{0.03}$\markc 	& 3.7$_{2.0}^{0.8}$\markc	& 4 	& ...  				& ... 	& ... 		& ... \\
CXOU J010043.1$-$721134		& 3.9 & $-$0.29		& 0.35(6)\markc			& 1.8(1)\markc			& 5 	& ...  				& ... 	& ... 		& ... \\
1E 1048.1$-$5937 		& 4.2 & $-$0.65 		& 0.56(1) 			& 3.14(11) 			& 6 	& ...  				& ... 	& ... 		& ... \\
1RXS J170849.0$-$400910		& 4.7 & $-$0.16 		& 0.456$_{0.007}^{0.004}$\markc	& 2.792$_{0.012}^{0.008}$\markc	& 7 	& 1.01(12) 			& 12 	& 1.44(45) 	& 12 \\
SGR 1900+14 			& 6.4 & $-$2.91 		& 0.47(2)\markc			& 1.9(1)\markc			& 8 	& ... 				& ... 	& 3.1(5) 	& 14 \\
1E 1841$-$045  			& 7.1 & $-$0.30 		& 0.44(2) 			& 2.0(3) 			& 9 	& 0.72(15) 			& 12 	& 1.32(11) 	& 12 \\
SGR 0526$-$66 			& 7.3 & $-$1.00 		& 0.48(5) 			& 3.12(8) 			& 10 	& ...  				& ... 	& ... 		& ... \\
SGR 1806$-$20 			& 21  & $-$9.62 		& 0.476(39)\markd		& 1.67(15) 			& 11 	& ... 				& ... 	& 2.0(2)\markc 	& 15 \\\hline
\end{tabular}}
\end{center}
\label{ta:gammas}
\vspace{-0.1in}
\footnotesize{
\tablenotetext{\rm a}{Numbers in parentheses represent 1$\sigma$ uncertainties in the least significant digit(s) unless otherwise indicated (see c).}
\tablenotetext{\rm b}{$B$ and $\dot{\nu}$ for this source are not yet measured precisely; see \citet{wkga09}.}
\tablenotetext{\rm c}{These uncertainties represent 90\% confidence intervals.}
\tablenotetext{\rm d}{The BB temperature is from a different observation as the reported $\Gamma_s$; see \citet{wkf+07}.}
}
\footnotesize{
Refs:
[1] \citet{wkt+04};
[2] \citet{icd+07};
[3] \citet{wae+96};
[4] \citet{gg07};
[5] \citet{mgr+05};
[6] \citet{tgd+08};
[7] \citet{rio+07};
[8] \citet{met+06};
[9] \citet{mskk03};
[10] \citet{kkm+03};
[11] \citet{wkf+07};
[12] \citet{khdc06};
[13] \citet{dkh+08};
[14] \citet{gmte06};
[15] \citet{emt+07}}
\vspace{-0.1in}
\end{table}

\clearpage

\begin{figure}
\plotone{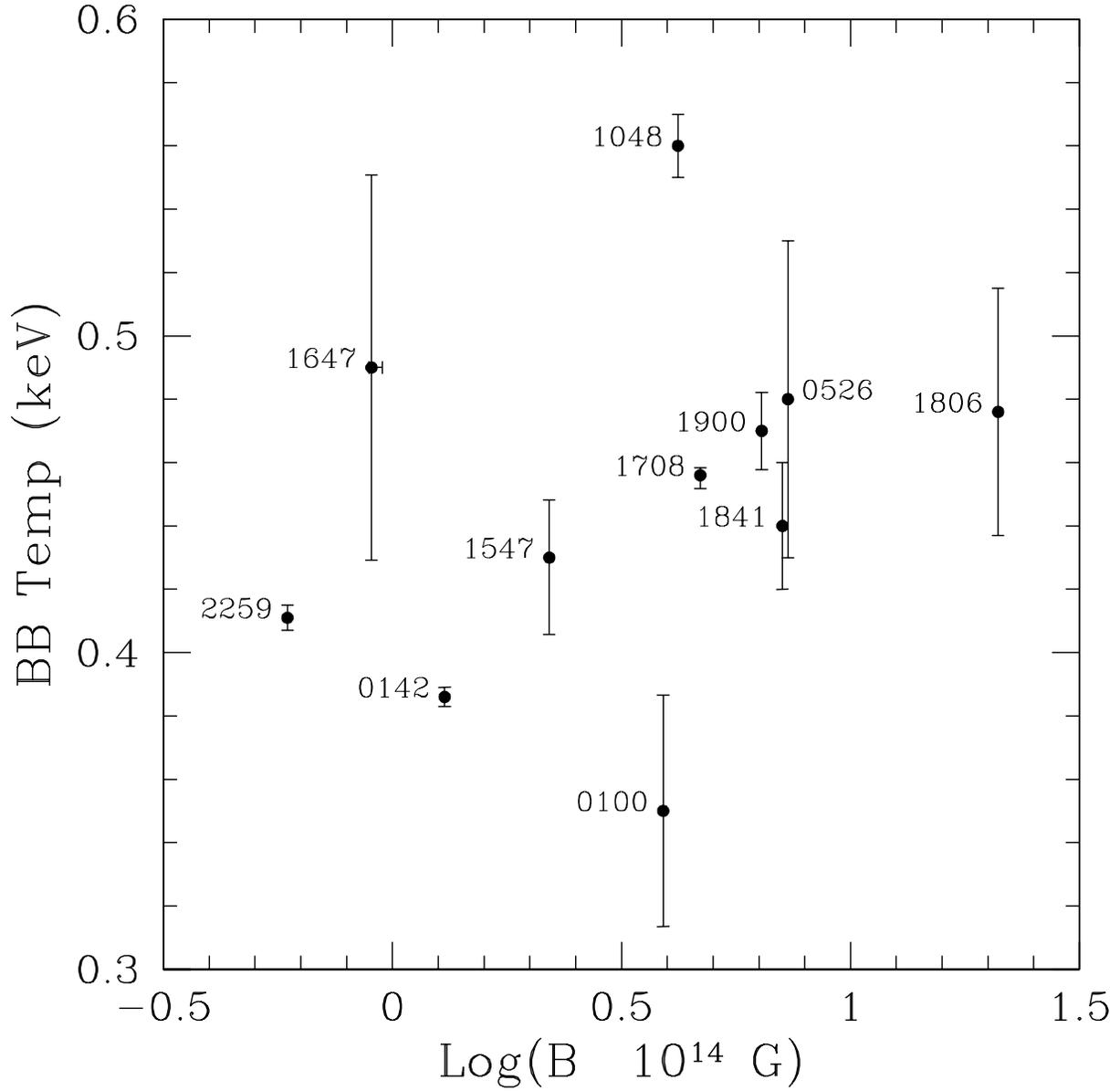}
\figcaption{Blackbody temperature
versus $B$, for all magnetars for which $\Gamma_s$ is
measured.  See text and Table~1 for data used and references.
Error bars represent 1$\sigma$ uncertainties.
\label{fig:bb}
}
\end{figure}

\begin{figure}
\plottwo{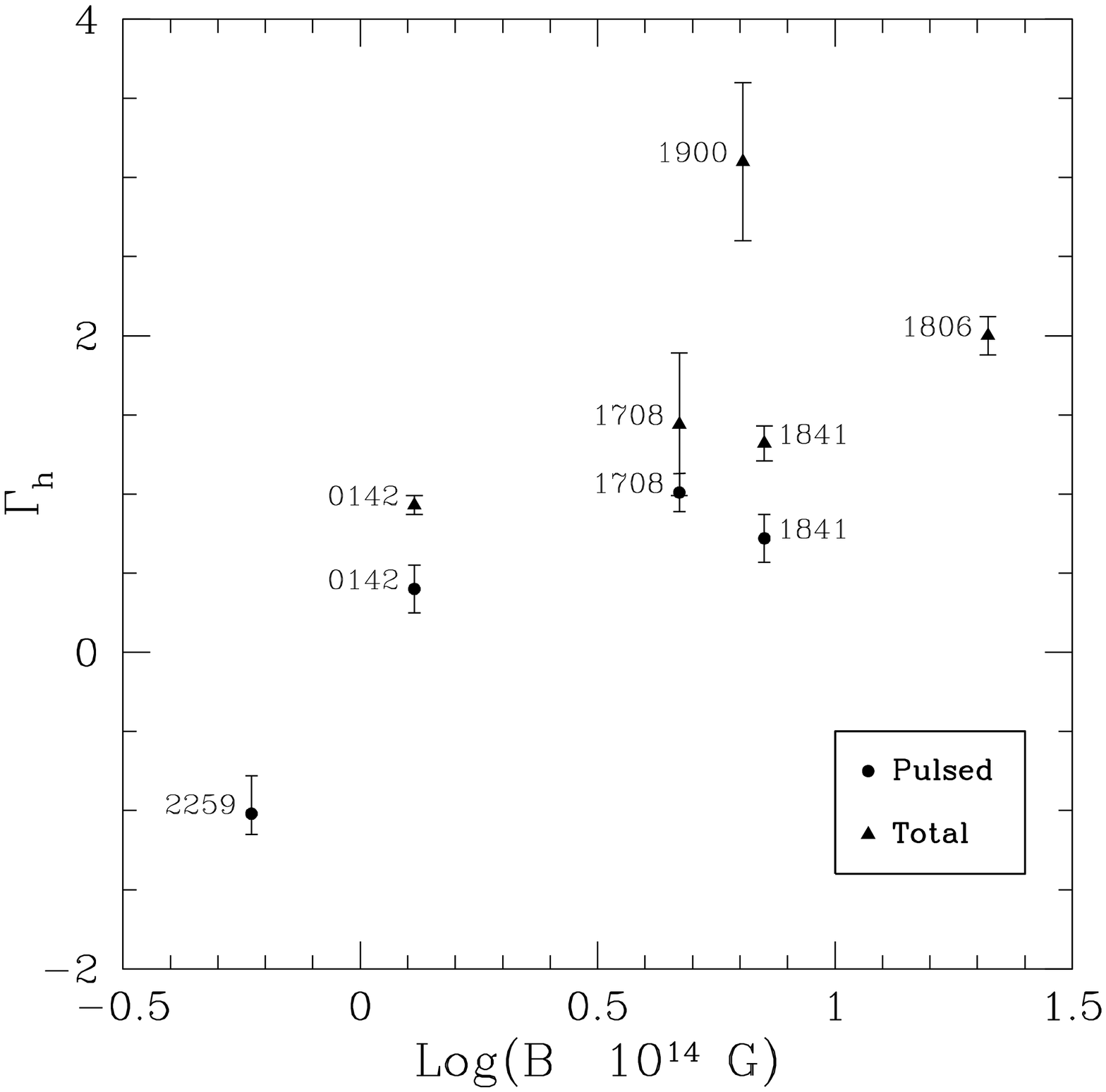}{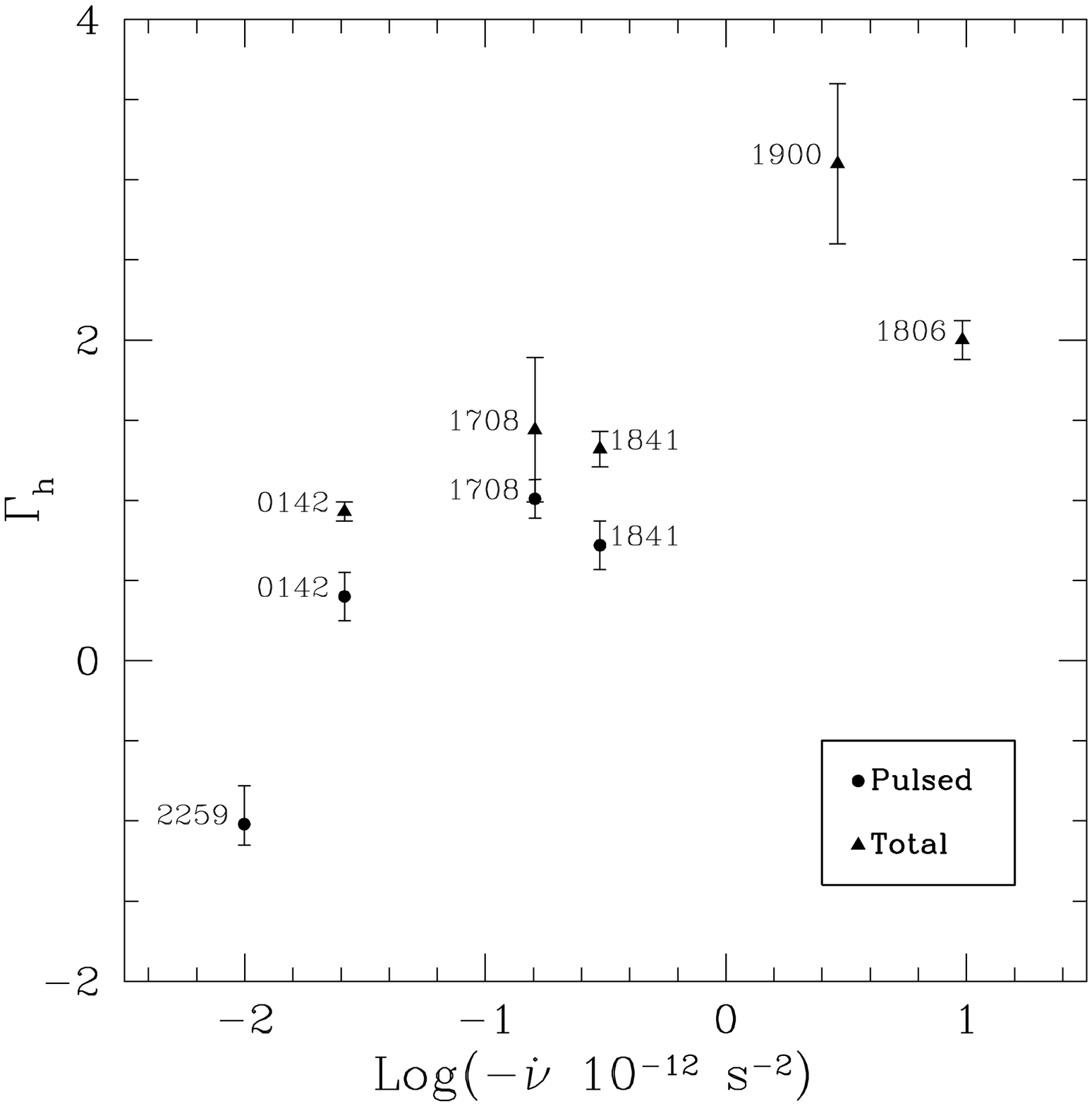}
\figcaption{Left:  $\Gamma_h$ versus $B$, for
all magnetars for which $\Gamma_h$ is measured.
See text and Table~1 for data used and references.  
Circles represent pulsed flux, triangles represent total flux.
Right:  Same but versus $\dot{\nu}$.
Both:  Error bars represent 1$\sigma$ uncertainties.
\label{fig:gammah}
}
\end{figure}

\begin{figure}
\plottwo{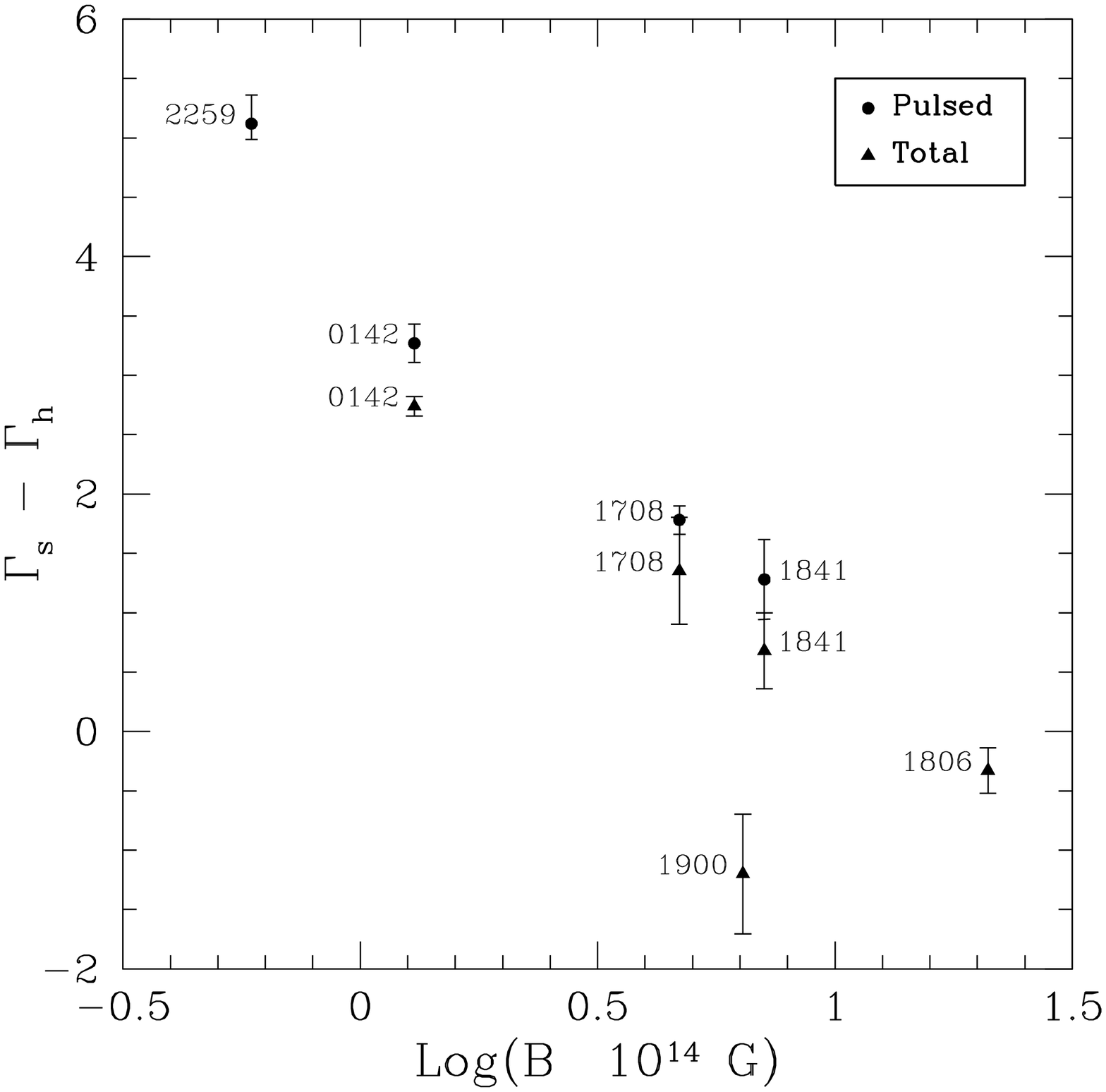}{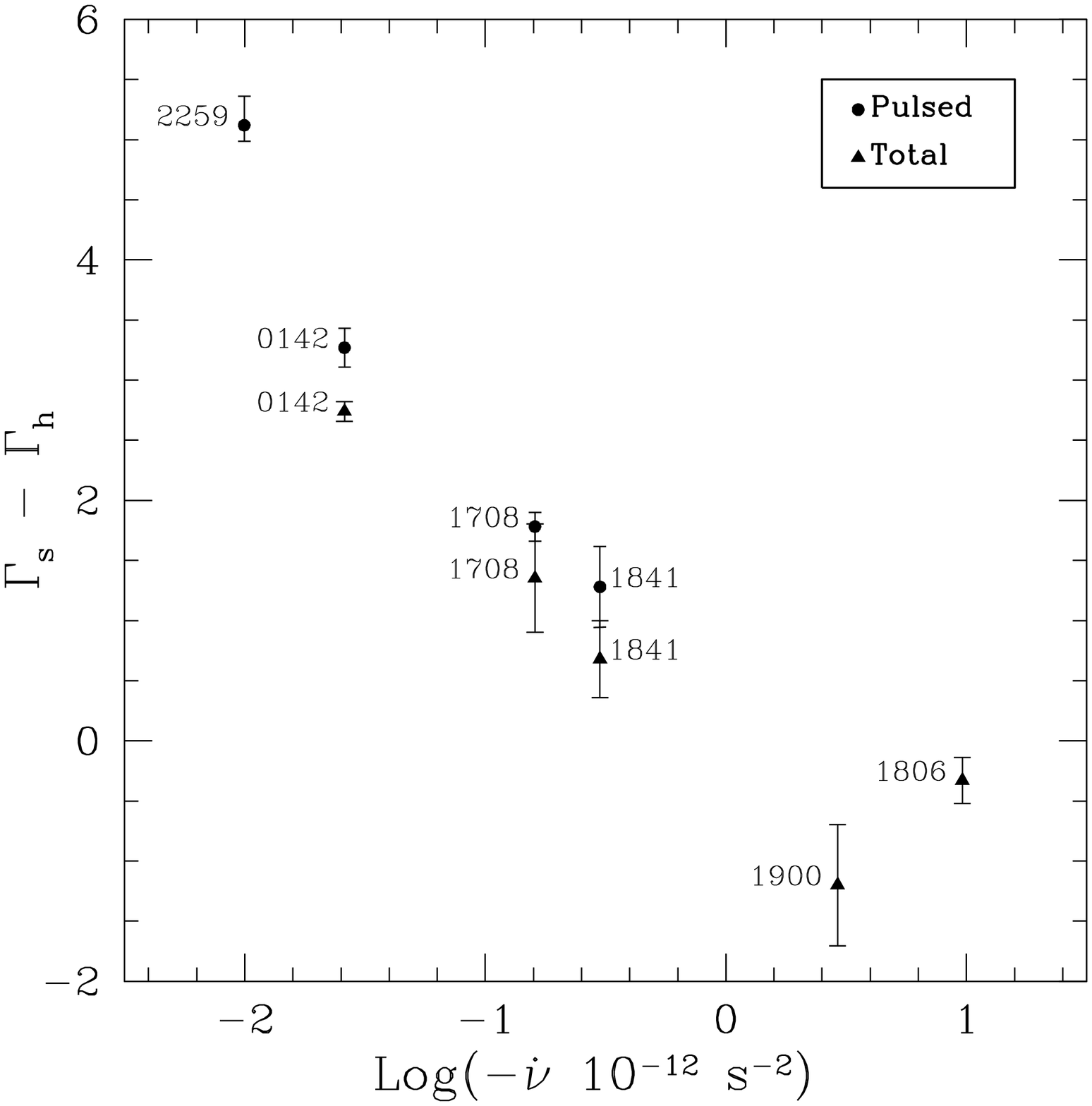}
\figcaption{Left:  Spectral turnover, $\Gamma_s - \Gamma_h$,
versus $B$, for all magnetars for which both $\Gamma_s$ and $\Gamma_h$ are
measured.  Circles represent pulsed flux, triangles represent total flux.  See text and Table~1 for data used and references.  Right:  Same but versus $\dot{\nu}$.
Both:  Error bars represent 1$\sigma$ uncertainties.  All known RPPs
for which the quantity has been measured
have $\Gamma_s - \Gamma_h \approx 0$, with $B$ in the range $1-10 \times 10^{12}$~G.
\label{fig:turnover}
}
\end{figure}

\end{document}